\documentclass[%
 reprint,
 amsmath,amssymb,
 aps,
prl,
superscriptaddress
]{revtex4-2}

\usepackage{graphicx}% Include figure files
\usepackage{dcolumn}% Align table columns on decimal point
\usepackage{bm}% bold math
\usepackage[colorlinks=true, allcolors=blue]{hyperref}
\usepackage{braket}
\usepackage{bbding}
\usepackage{natbib}
\usepackage{gensymb}
\usepackage{balance}
\begin{document}

\preprint{APS/123}

\title{Nonreciprocal magnons in layered antiferromagnets $\mbox{VP}X_{3}$ ($X$\,$=$\,$\mbox{S}$,\, $\mbox{Se}$,\,$\mbox{Te}$)} %: \\
%External-field Manipulation and Layer Dependence}% Force line breaks with %\\
%\thanks{A footnote to the article title}%

% \altaffiliation{SPMS, Nanyang Techonology University.}%Lines break automatically or can be forced with \\
\author{Quanchao Du}
\email{These authors contributed equally to this work.}
% \altaffiliation{SPMS, Nanyang Techonology University.}%Lines break automatically or can be forced with \\
\affiliation{%
 Ministry of Education Key Laboratory for Nonequilibrium Synthesis and Modulation of Condensed Matter, Shaanxi Province Key Laboratory of Advanced Functional Materials and Mesoscopic Physics, School of Physics, Xi’an Jiaotong University, Xi’an 710049, China %\textbackslash\textbackslash
}%  

\author{Zhenlong Zhang}
\email{These authors contributed equally to this work.}
% \altaffiliation{SPMS, Nanyang Techonology University.}%Lines break automatically or can be forced with \\
\affiliation{%
 Ministry of Education Key Laboratory for Nonequilibrium Synthesis and Modulation of Condensed Matter, Shaanxi Province Key Laboratory of Advanced Functional Materials and Mesoscopic Physics, School of Physics, Xi’an Jiaotong University, Xi’an 710049, China %\textbackslash\textbackslash
}%

\author{Jinyang Ni}
\email{jyni@xjtu.edu.cn}
\affiliation{%
 Ministry of Education Key Laboratory for Nonequilibrium Synthesis and Modulation of Condensed Matter, Shaanxi Province Key Laboratory of Advanced Functional Materials and Mesoscopic Physics, School of Physics, Xi’an Jiaotong University, Xi’an 710049, China %\textbackslash\textbackslash
}% 
\affiliation{Key Laboratory of Computational Physical Sciences (Ministry of Education), Institute of Computational Physical Sciences, State Key Laboratory of Surface Physics and Department of Physics, Fudan University, Shanghai 200433, China}

\author{Zhijun Jiang}
\email{zjjiang@xjtu.edu.cn}
% \altaffiliation{SPMS, Nanyang Techonology University.}%Lines break automatically or can be forced with \\
\affiliation{%
 Ministry of Education Key Laboratory for Nonequilibrium Synthesis and Modulation of Condensed Matter, Shaanxi Province Key Laboratory of Advanced Functional Materials and Mesoscopic Physics, School of Physics, Xi’an Jiaotong University, Xi’an 710049, China %\textbackslash\textbackslash
}%  
\affiliation{Key Laboratory of Computational Physical Sciences (Ministry of Education), Institute of Computational Physical Sciences, State Key Laboratory of Surface Physics and Department of Physics, Fudan University, Shanghai 200433, China}
\author{Laurent Bellaiche}
\affiliation{%
Smart Ferroic Materials Center, Department of Physics and Institute for Nanoscience and Engineering, University of Arkansas, Fayetteville, Arkansas 72701, USA
}%
\affiliation{%
Department of Materials Science and Engineering, Tel Aviv University, Ramat Aviv, Tel Aviv 6997801, Israel
}%

%\collaboration{MUSO Collaboration}%\noaffiliation

%\author{Charlie Author}
 %\homepage{http://www.Second.institution.edu/~Charlie.Author}
%\affiliation{
% Second institution and/or address\\
% This line break forced% with \\
%}%
%\affiliation{
% Third institution, the second for Charlie Author
%}%
%\author{Delta Author}
%\affiliation{%
% Authors' institution and/or address\\
% This line break forced with \textbackslash\textbackslash
%}%

%\collaboration{CLEO Collaboration}%\noaffiliation

%\date{\today}% It is always \today, today,
             %  but any date may be explicitly specified

\begin{abstract}
Nonreciprocal magnons, characterized by propagation with differing energies along the $k$ and $-k$ directions, are crucial for modern spintronics applications. However, their realization in van der Waals layered antiferromagnets remains elusive. In this letter, we report robust nonreciprocal magnon behavior in layered honeycomb antiferromagnets $\mbox{VP}{X}_{3}$ ($X = \mbox{S}, \mbox{Se}, \mbox{Te}$). Our results demonstrate that, in addition to their intrinsic Dzyaloshinskii-Moriya interaction (DMI), the nonreciprocity of magnons is strongly influenced by the layer number, interlayer coupling, and magnon-magnon interactions. More importantly, in such layered antiferromagnets, the magnon nonreciprocity exhibits an asymmetric periodic dependence on the N\'{e}el vector, offering a novel route for experimentally probing antiferromagnetic order parameters in the 2D limit.
%\begin{description}
%\item[Usage]
%xxx.
%\item[Structure]
%xxx. 
%\end{description}
\end{abstract}

%\keywords{Suggested keywords}%Use showkeys class option if keyword
                              %display desired
\maketitle

%\tableofcontents

%\section{Introduction}
\textit{Introduction.} 
In condensed matter physics, the nonreciprocal properties of particles or quasiparticles refer to the asymmetric behavior of the band dispersion or transport characteristics along opposite directions\,\cite{feng2011nonreciprocal,ramezani2018nonreciprocal,xu2020nonreciprocal,caloz2018electromagnetic,cheong2018broken,wang2019nonreciprocity,wakatsuki2018nonreciprocal,nagaosa2024nonreciprocal,wakatsuki2017nonreciprocal,tokura2018nonreciprocal,itahashi2020nonreciprocal}. This nonreciprocal behavior for electrons is widely observed in noncentrosymmetric systems with pronounced spin–orbit coupling effect, such as the Rashba or Dresselhaus spin splitting\,\cite{rashba1960properties,bychkov1984properties,xiao2012coupled}, which leads to the direction-selective electronic transports. Intriguingly, the direction-dependent nature offers opportunities for manipulating transport properties through electric or optical fields, thereby providing a foundational platform for the development of novel functional spintronic\,\cite{cornelissen2015long,zhang2012magnon,li2016observation,bedoya94competing,cornelissen2016magnon,lebrun2018tunable,han2020birefringence,lebrun2020long,schlitz2021control,zhang2014electric}.

Magnons, as the quanta of collective spin excitations in ordered magnets, are also known to exhibit nonreciprocal behavior\,\cite{hayami2016asymmetric,gitgeatpong2017nonreciprocal,iguchi2015nonreciprocal,costa2020nonreciprocal,ni2025nonvolatile,sato2019nonreciprocal,guckelhorn2023observation, onose_science_2010_329}. The primary driving forces behind nonreciprocal magnons are the asymmetric spin exchange interactions\,\cite{hayami2016asymmetric, hayami2022essential, matsumoto2020nonreciprocal}, including the bond-dependent spin interaction and Dzyaloshinskii-Moriya interactions(DMI)\,\cite{dzyaloshinsky_JPCS_1958_4, moriya_PR_1960_120}. This typically requires material candidates to possess a strong spin orbital coupling effect (SOC) and break inversion symmetry in spin space. However, realizing robust nonreciprocal magnons in 2D van der Waals (vdW) antiferromagnetic (AFM) insulators remains elusive. For example, in transition metal phosphorus trichalcogenides
$M\mbox{P}X_{3}$ ($M = \mbox{Fe}, \mbox{Co}, \mbox{Ni}$, $X = \mbox{S}, \mbox{Se}, \mbox{Te}$)\,\cite{wang2018new, ni2021direct, kim2019suppression, klaproth2023origin, ni2021imaging, liu2023probing, mak2019probing}, they generally adopt a zigzag antimagnetic ground state that preserves inversion symmetry, inherently forbidding non-reciprocal magnons. Moreover, while the honeycomb AFM insulators with N\'{e}el order, such as $\mbox{MnPS}_{3}$ or $\mbox{MnPSe}_{3}$\,\cite{pich1995spin,cheng2016spin,wildes2021search,matsuoka2024mpx}, permit the emergence of nonreciprocal magnons, their extremely weak DMI and XY-type spin order pose challenges for experimental verification. Therefore, the search for robust nonreciprocal magnons 2D layered antiferromagnets is essential, both for fundamental physics and future wave-based applications\,\cite{chumak2015magnon, baltz2018antiferromagnetic, jungwirth2016antiferromagnetic, li2020spin}.

In this Letter, we report the robust nonreciprocal magnons in layered AFM insulators $\mbox{VP}X_{3}$, most notably in $\mbox{VPSe}_{3}$ and $\mbox{VPTe}_{3}$. By performing density-functional-theory (DFT) calculations and effective spin model analysis, we demonstrate that the nonreciprocal magnons arise from the giant second-nearest-neighbor (2NN) DMI between $\mbox{V}^{2+}$ ions, and are strongly coupled to their N\'{e}el vector. In addition to the giant nonreciprocal magnons observed in the monolayer, multilayer $\mbox{VP}X_{3}$ exhibits intriguing odd-even layer dependence in nonreciprocal magnon behavior. This intricate coupling allows for the control of nonreciprocal magnons through external pressure or magnetic fields, thereby offering a high degree of tunability. 

\begin{figure*}
\includegraphics[width=0.85\textwidth]{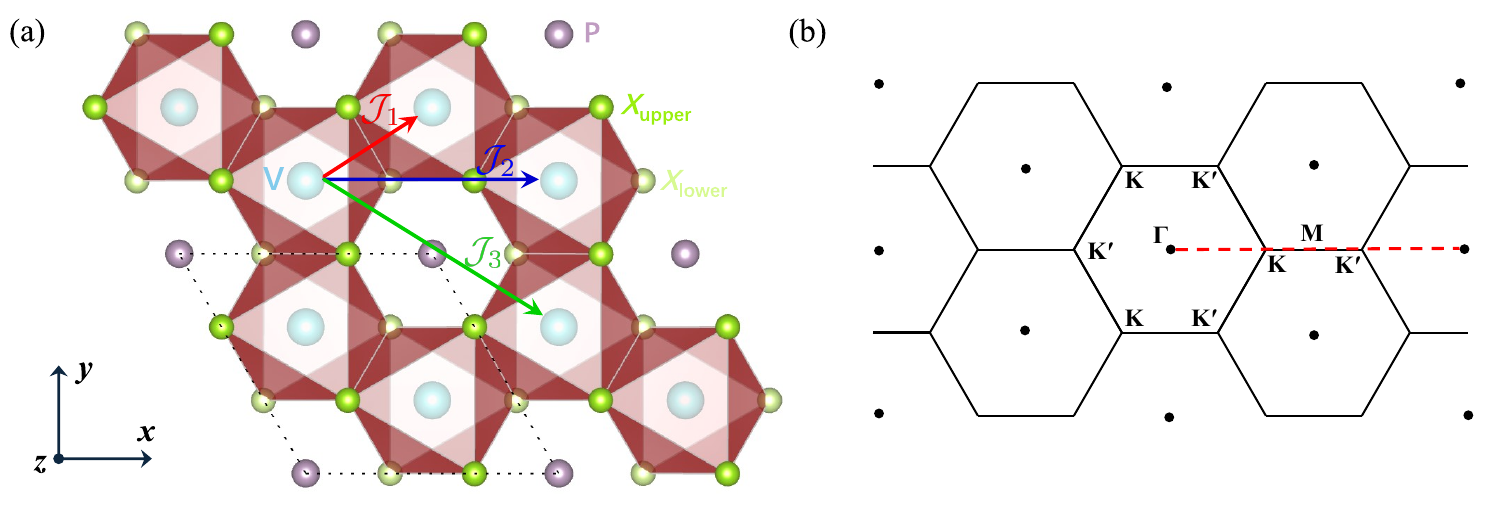}% Here is how to import EPS art
\caption{\label{fig1}(a) Top view of geometric structure of monolayer $\mbox{VP}X_{3}$. The NN, 2NN and 3NN spin exchange paths are labeled by red ($\boldsymbol{\delta_{i}}$), blue ($\boldsymbol{\mu_{i}}$) and green ($\boldsymbol{\zeta_{i}}$) vectors, respectively. (b) Schematic showing a plane in reciprocal space for
the honeycomb lattice with N\'{e}el order. A Brillouin zone center is marked as $\Gamma$, and high-symmetry points on the Brillouin zone boundary are marked with $\mbox{K}$, $\mbox{M}$, and $\mbox{K}^{\prime}$. }
\end{figure*} 

\textit{Symmetry and ground states of $\mbox{VP}X_{3}$.} 
Layered AFM insulators $\mbox{VP}X_{3}$ typically crystallize in $P\bar{3}m1$ symmetry\,\cite{wang2018new, liu2023probing}, where each layer can be viewed as honeycomb lattice of $\mbox{V}^{2+}$ ions coordinated by distorted octahedra, with adjacent layers separated by the vdW interactions, as illustrated in Fig.\,\ref{fig1}(a). Our DFT calculations show that the localized $\mbox{V}^{2+}$ magnetic ion adopts a high spin $S$\,$=$\,$3/2$ configuration, as described in the Supplemental Material\,\cite{Supplemental_Materials}\,(see also Refs.\,\cite{blochl1994projector,kresse1996efficiency,perdew1996generalized,liechtenstein1995density,grimme2010consistent,xiang2011predicting,xiang2013magnetic,cococcioni2005linear,timrov2022hp,liechtenstein1987local,grotheer2001fast,mostofi2008wannier90,he2021tb2j,kitaev2006anyons,xu2018interplay}\,therein), forming a honeycomb lattice in $\mbox{VP}X_{3}$ with an easy-axis N\'{e}el order as magnetic ground state. As confirmed by the DFT calculations presented in
Tab.\,\ref{tab_1} and Tab.\,S1\,\cite{Supplemental_Materials}, this Ising-type N\'{e}el order emerges from the robust first-nearest-neighbor (NN) AFM Heisenberg spin exchange and out-of-plane single ion anisotropy (SIA) of localized $\mbox{V}^{2+}$ moments. Furthermore, the $z$ component DMI interactions among second-nearest-neighbor (2NN) $\mbox{V}^{2+}$ ions in $\mbox{VPSe}_{3}$ and $\mbox{VPTe}_{3}$ are significant, with magnitudes of $0.1\,\mbox{meV}$ and $0.3\,\mbox{meV}$, respectively. Note that the NN DMI is absent due to the inversion symmetry. Given the weak SOC of $\mbox{V}^{2+}$ ions with high spin configuration\,\cite{khomskii2014transition}, the significant SIA and 2NN DMI of $\mbox{VPSe}_{3}$ and $\mbox{VPTe}_{3}$ are mainly attributed to the heavy ligands $\mbox{Se}$ or $\mbox{Te}$. Overall, the robust honeycomb easy-axis N\'{e}el order, combined with the significant DMI, makes $\mbox{VPX}_{3}$ an ideal candidate to realize nonreciprocal magnons.

\textit{Monolayer $\mbox{VP}X_{3}$.} We first examine the nonreciprocal magnons of monolayer $\mbox{VP}{X}_{3}$, where the spin model, including spin exchange interactions up to the third-nearest-neighbor (3NN), can be expressed as
\begin{equation} \label{Ham_1}
\begin{split}
{\cal \hat{H}}_{1}  = & {\cal J}_{1} \sum_{\langle {i,j} \rangle}{{\cal S}_{i} \cdot{\cal S}_{j}} + {\cal J}_{2} \sum_{\langle {i,k} \rangle}{{\cal S}_{i} \cdot{\cal S}_{j}} + {\cal J}_{3} \sum_{\langle {i,l} \rangle}{{\cal S}_{i} \cdot{\cal S}_{j}} \\ & + {\cal D}_{z} \sum_{ \langle {i,k} \rangle} \boldsymbol{\nu}_{ij}{ \cdot \left( {\cal S}_{i} \times {\cal S}_{j} \right)}  + \sum_{i} {\cal K} \left( {\cal S}^{z}_{i} \right)^{2} . 
\end{split}
\end{equation}
The first to third terms of this model correspond to the Heisenberg exchange interactions among NN, 2NN and 3NN $\mbox{V}^{+2}$ spin moments on the honeycomb lattice, respectively. The fourth term is 2NN $z$-component DMI ${\cal D}_{z}$\,\cite{dzyaloshinsky_JPCS_1958_4, moriya_PR_1960_120, mook_PRX_2021_11}, where the direction is defined as $\boldsymbol{\nu}_{ij}= 2\sqrt{3}\boldsymbol{\delta}_{i}\times\boldsymbol{\delta}_{j} = \pm{\boldsymbol{z}}$. The last term represents the easy-axis SIA with ${\cal K}<0$. In the absence of 2NN DMI, our ${\cal J}$-${\cal D}$-${\cal K}$ spin model is evidently invariant under the combined symmetry time-reversal ${\cal T}$ and $\pi$ rotation along $x$-axis $C_{2x}$ in the spin space. The combined symmetry $C_{2x}\cal{T}$ is known as the effective time-reversal symmetry ${\cal T}^{\prime}$ for magnons, which can be broken by 2NN ${\cal D}_{z}$\,\cite{cheng2016spin, mook_PRX_2021_11, ni2024magnon}. On the other hand, in the monolayer honeycomb lattice with N\'{e}el order, the inversion symmetry ${\cal P}$ is naturally broken\,\cite{brinkman1966theory}. 

\begin{table}[b]
\caption{\label{tab_1}Calculated spin exchange parameters (with units of meV)
of monolayer $\mbox{VP}{X}_{3}$.}
\begin{ruledtabular}
\begin{tabular}{cccccc}
$Material$  & ${\cal J}_{1}$  & \multicolumn{1}{c}{${\cal D}_{x}$} & \multicolumn{1}{c}{${\cal D}_{y}$}  & \multicolumn{1}{c}{${\cal D}_{z}$}  & \multicolumn{1}{c}{${\cal K}$} \tabularnewline
\hline 
$\mbox{VPS}_{3}$  & 23.67  & $-$0.006  & $-$0.012 & $-$0.03 & $-$0.02 \tabularnewline
$\mbox{VPSe}_{3}$  & 15.73  & $-$0.026  & $-$0.045 & $-$0.11 & $-$0.14 \tabularnewline
$\mbox{VPTe}_{3}$  & 7.73  & $-$0.09  & $-$0.13 & $-$0.31 & $-$0.41 \tabularnewline
\end{tabular}
\end{ruledtabular}

\end{table} 

The linear spin wave (LSW) model in Eq.\,(\ref{Ham_1}) can be solved by employing the Holstein-Primakoff (HP) transformation\,\cite{holstein1940field}, ${\cal S}^{+}_{i,\uparrow}$\,$\approx$\,$\sqrt{2S}\hat{a}_{i}$, ${\cal S}^{+}_{i,\downarrow}$\,$\approx$\,$\sqrt{2S}\hat{b}^{\dagger}_{i}$, ${\cal S}^{z}_{i,\uparrow}$\,$=$\,$S$\,$-$\,$\hat{a}_{i}^{\dagger}\hat{a}_{i}$ and ${\cal S}^{z}_{i,\downarrow}$\,$=$\,$\hat{b}_{i}^{\dagger}\hat{b}_{i}$\,$-$\,$S$, with $\hat{a}_{i}^{\dagger}$\,$(\hat{b}_{i}^{\dagger})$ creating a magnon on A(B) sublattice in the $i$-th unit cell. After Fourier transformation, the Hamiltonian can be expressed in the spinor basis $\psi^{\dagger}_{k} = (\hat{a}_{-k}, \hat{b}_{-k}, \hat{a}^{\dagger}_{k}, \hat{b}^{\dagger}_{k})$ as ${\cal \hat{H}}_{1} = \sum_{k}\psi^{\dagger}_{k} {\cal \hat{H}}_{1k} \psi_{k}$. Neglecting the zero-point energy, ${\cal \hat{H}}_{1k}$ reads as, 
%\begin{widetext}
\begin{equation} {\label{Ham_k}}
 {\cal \hat{H}}_{1k}/S =  \left(\lambda + {g_{k}}\right) I + \left(\begin{matrix}
    {f_{k}}  &0 & 0 & \gamma_{k}\\
   0& {f_{k}} & \gamma_{k}^{\dagger} & 0 \\
    0 & \gamma_{k} & -{f_{k}} &  0 \\
    \gamma^{\dagger}_{k} & 0 &  0 & -{f_{k}}
\end{matrix} \right),
\end{equation}
%\end{widetext}}
where $\lambda = 3{\cal J}_{1}+6{\cal J}_{2}+3{\cal J}_{3} - 2{\cal K}$, $\gamma_{k} =  {\cal J}_{1}\sum_{i} \mbox{exp}(i \textbf{\textit{k}}\cdot \boldsymbol{\delta}_{i}) + {\cal J}_{3}\sum_{i} \mbox{exp}(i \textbf{\textit{k}}\cdot \boldsymbol{\zeta}_{i})$, $f_{k} = {\cal D}_{z}\sum_{i\in odd}2\mbox{sin}(\textbf{\textit{k}}\cdot \boldsymbol{\mu}_{i})$ and $g_{k} = {\cal J}_{2}\sum_{i\in odd}2\mbox{cos}(\textbf{\textit{k}}\cdot \boldsymbol{\mu}_{i})-6$. Here, $\boldsymbol{\delta_{i}}$, $\boldsymbol{\mu}_{i}$ and $\boldsymbol{\zeta}_{i}$ represent the vectors linking NN, 2NN and 3NN sites of honeycomb lattice, respectively, as illustrated in Fig.\,\ref{fig1}(a). 

\begin{figure}
\includegraphics[width=0.475\textwidth]{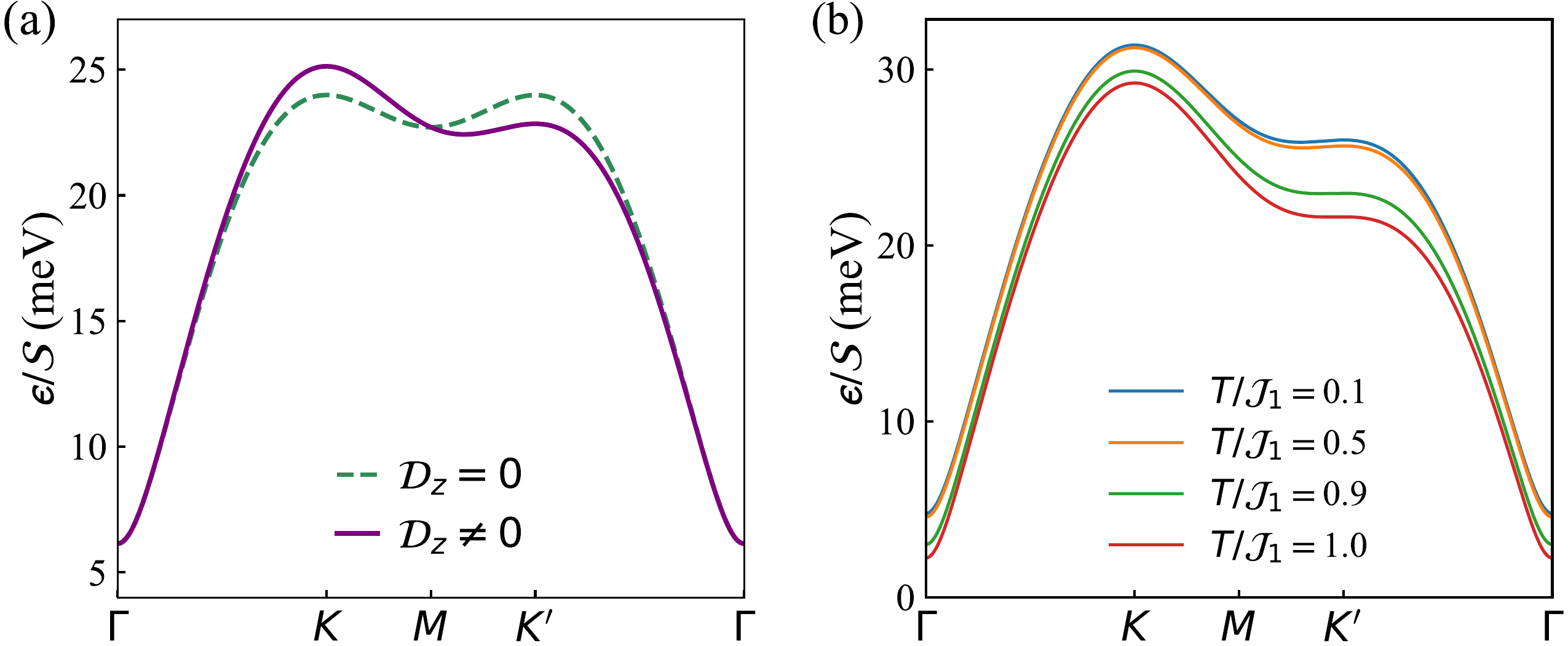}% Here is how to import EPS art
\caption{\label{fig2}(a) The magnon bands of monolayer $\mbox{VPTe}_{3}$, where dotted line represents the case without ${\cal D}_{z}$, and solid line correspondes to the case including ${\cal D}_{z}$. (b) The renormalized magnon bands of monolayer $\mbox{VPTe}_{3}$ under different temperatures $T$.}
\end{figure}

To diagonalize Eq.\,(\ref{Ham_k}), we employ the Bogoliubov transformation for quadratic bosonic Hamiltonian\,\cite{valatin1958comments, bogoljubov1958new}. Under the Bogoliubov transformation, the transformed spinor basis can be expressed as
\begin{equation}
    \Psi^{\dagger}_{k} = {\cal R}\psi_{k} = (\hat{\alpha}_{-k}, \hat{\beta}_{-k}, \hat{\alpha}^{\dagger}_{k}, \hat{\beta}^{\dagger}_{k}), 
\end{equation}
where matrix ${\cal R}$ satisfies the para-unitary relation, ${\cal R}\eta {\cal R}^{\dagger} = \eta$, with $\eta = \sigma_{z} \otimes I$. This feature ensures that ${\cal R} {\cal \hat{H}}_{1k} {\cal R}^{\dagger}$ becomes diagonal, with its diagonal elements corresponding to the magnon eigenmodes, expressed as 
\begin{equation}\label{eigen_AFM}
\begin{split}
        {{\epsilon}}_{\alpha, \pm} &= \pm \sqrt{\left(\lambda + g_{k}\right)^{2}-|\gamma_{k}|^{2}} + f_{k},\\
        {{\epsilon}}_{\beta, \pm} &= \pm \sqrt{\left(\lambda + g_{k}\right)^{2}-|\gamma_{k}|^{2}} + f_{k}. 
\end{split}
\end{equation}
Clearly, there are two redundant solutions arising from the particle-hole symmetry of Eq.\,(\ref{Ham_k}), which can be disregarded. Two physical solutions, $\epsilon_{\alpha}$ and $\epsilon_{\beta}$, associated with the corresponding eigenkets $\ket{\hat{\alpha}^{\dagger}_{k}}$ and $\ket{\hat{\beta}^{\dagger}_{k}}$, are two-fold degenerate and differed in spin chirality. Since our ${\cal J}$-${\cal D}$-${\cal K}$ spin model preserves $C_{3z}$ rotational symmetry, the $z$-component of the total spin, ${\cal S}^{z}$\,$=$\,$\sum_{i}{\cal S}^{z}_{i,\uparrow}$\,$+$\,${\cal S}^{z}_{i,\downarrow}$, serves as a good quantum number for describing
magnon Hamiltonian. This can be verified by the HP transformation, where ${\cal S}^{z}$\,$=$\,$\sum_{k}{\cal S}^{z}_{k}$\,$=$\,$\sum_{k}-\hat{a}^{\dagger}_{k}\hat{a}_{k}$\,$+$\,$\hat{b}^{\dagger}_{k}\hat{b}_{k}$ and ${\cal S}^{z}_{k}$ commutes with ${\hat{\cal H}_{1k}}$. By invoking the Bogoliubov  transformation, we further obtain ${\cal S}^{z}$\,$=$\,$\sum_{k}$\,$-$\,$\hat{\alpha}^{\dagger}_{k}\hat{\alpha}_{k}$\,$+$\,$ \hat{\beta}^{\dagger}_{k}\hat{\beta}_{k}$, thus, $\bra{\alpha_{k}}{\cal S}^{z}\ket{\alpha^{\dagger}_{k}}$\,$=$\,$-1$ and $\bra{\beta_{k}}{\cal S}^{z}\ket{\beta^{\dagger}_{k}}$\,$=$\,$1$. This indicates that $\alpha$ and  $\beta$ magnon mode carry opposite spin angular momentum along the $z$ direction $\langle{{\cal S}^{z}}\rangle$ with chirality $\mp{1}$.

The magnons bands of monolayer $\mbox{VPTe}_{3}$ are presented in Fig.\,\ref{fig2}(a), demonstrating that ${\cal K}$ can open a magnon gap at $\Gamma$ point. This feature disrupts the Goldstone-mode and ensures the long-range order of monolayer $\mbox{VP}X_{3}$ in the 2D limit\,\cite{mermin1966absence}. Obviously, the Heisenberg exchange interactions only affect the magnitude of the band dispersion but can not contribute to a magnon gap or the nonreciprocal magnons. Since the $f_{k}$ is odd with respect to $\boldsymbol{k}$, ${\cal D}_{z}$ induces an inequivalence in the magnon bands along $\Gamma$ to $\mbox{K}$ and $\mbox{K}^{\prime}$ to $\Gamma$ paths, resulting in nonreciprocity. For simplicity, the nonreciprocity of magnon bands can be defined as the energy difference between the $\mbox{K}$-points, $\varepsilon_{r}$\,$=$\,$ \epsilon(\mbox{K})$\,$-$\,$\epsilon(\mbox{K}^{\prime})$\,=\,$-6\sqrt{3}\ {\cal D}_{z}$. As shown in Fig.\,\ref{fig2}(b), $\varepsilon_{r}$ of the monolayer $\mbox{VPTe}_{3}$ is significant, reaching  $3.2\ \mbox{meV}$. Although the $\varepsilon_{r}$ of the monolayer $\mbox{VPSe}_{3}$ and $\mbox{VPS}_{3}$ is weaken than that of $\mbox{VPTe}_{3}$, it still reaches $2\ \mbox{meV}$ and $0.4\ \mbox{meV}$, respectively, which is sufficient for experimental detection. To verify the stability of nonreciprocal magnons under temperature variations, we include the magnon-magnon interactions in our analysis. As shown in Fig.\,\ref{fig2}(b), compared to LSW approach, the introduction of four-magnon interactions leads to enhanced magnon excitation at elevated temperatures\,\cite{rezende2019introduction, mook_PRX_2021_11, sourounis2024impact, paischer2024correlations}, while preserving its overall nonreciprocity. Importantly, the strength of its nonreciprocity can be enhanced at higher temperature regime.

It is useful to compare the nonreciprocal magnons between monolayer $\mbox{VP}X_{3}$ and monolayer $\mbox{MnP}X_{3}$, as they both adopt N\'{e}el ground state\,\cite{ni2021direct, wildes2021search}. Since magnetic anisotropy of $\mbox{MnP}X_{3}$ prefers the easy plane, their ${\cal T}^{\prime}$ changes as ${\cal C}_{2z}{\cal T}$, which can broken by the in-plane DMI. Our DFT calculations show that 2NN ${\cal D}_{x}$ or ${\cal D}_{y}$ of monolayer $\mbox{MnPSe}_{3}$ is weak with value of $0.01\ \mbox{meV}$, which is much smaller than in $\mbox{VPSe}_{3}$. The weak 2NN DMI makes it challenging to observe nonreciprocal magnons in 2D $\mbox{MnP}X_{3}$\,\cite{wildes2021search}.

\begin{figure}
\includegraphics[width=0.425\textwidth]{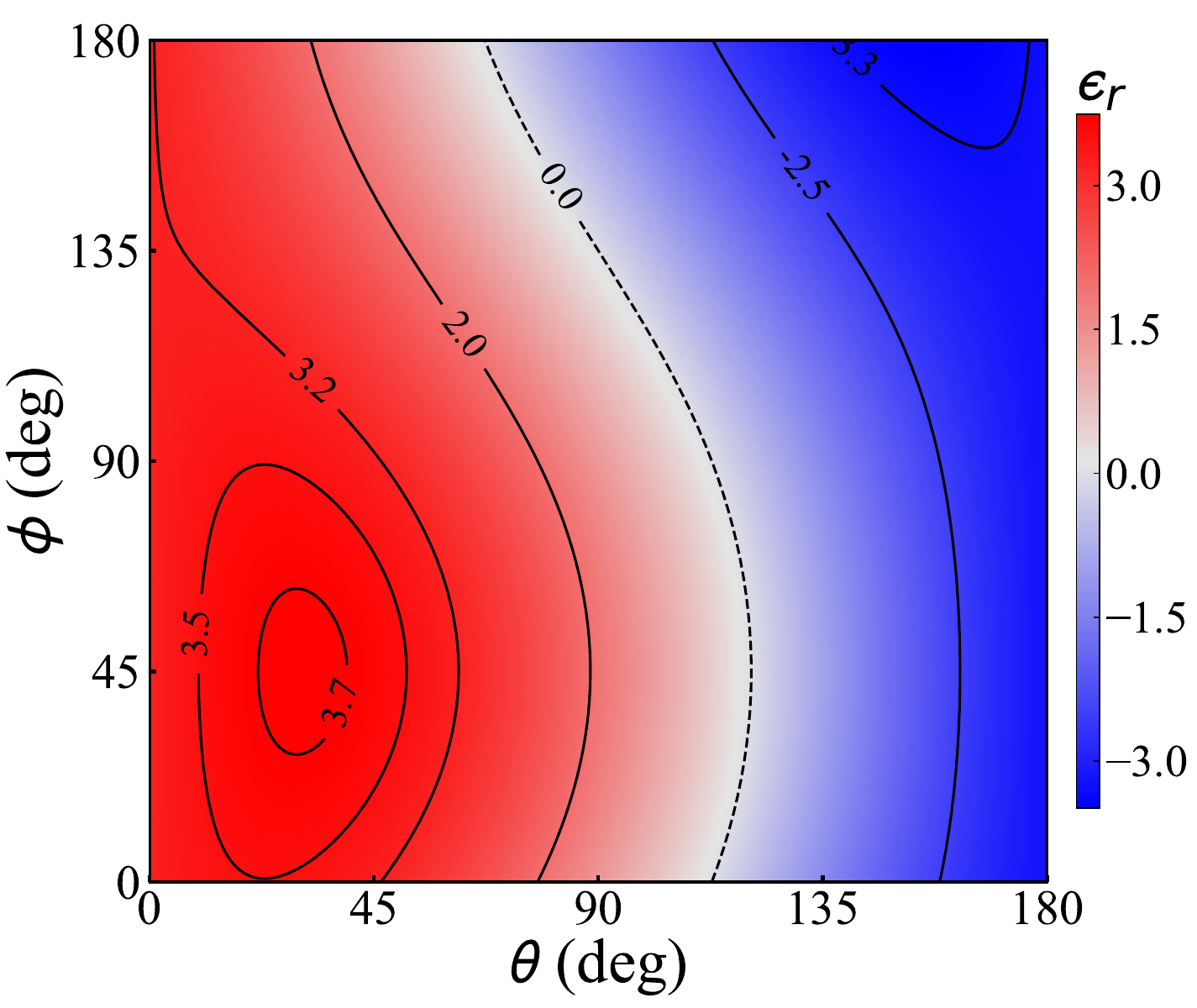}% Here is how to import EPS art
\caption{\label{fig3} The N\'{e}el vector dependence on magnon nonreciprocity in monolayer $\mbox{VPTe}_{3}$.}
\end{figure}

Notably, as illustrated in Tab.\,\ref{tab_1}, the in-plane components of 2NN DMI, ${\cal D}_{x}$ or ${\cal D}_{y}$, are also substantial in monolayer $\mbox{VPTe}_{3}$. This indicates a distinct dependence of ${\varepsilon}_{r}$ on N\'{e}el vector. By combining the LSW approach with Eq.\,(\ref{eigen_AFM}), the general relationship between N\'{e}el vector and ${\varepsilon}_{r}$ can be expressed as 
\begin{equation}
     \varepsilon_{r} = 6\sqrt{3}\left({\cal D}_{x}\mbox{sin}\theta\mbox{cos}\phi + {\cal D}_{y}\mbox{sin}\theta\mbox{sin}\phi + {\cal D}_{z}\mbox{cos}\theta\right), 
\end{equation}
where $\theta$ and $\phi$ are the angles with respect to the $z$ axis and $x$ axis, respectively. Clearly, it exhibits an asymmetric periodic dependence arising from the presence of in-plane DMI. For instance, when $\phi$\,$=$\,$0$, so that N\'{e}el vector lies in the $yz$ plane, $\varepsilon_{r}$ vanishes at $\theta$\,$\approx$\,$115^{\degree}$. As $\theta$ increases, $\varepsilon_{r}$ changes sign from positive to negative, and its magnitude differs significantly between $\theta$\,$=$\,$\mbox{0}$ and $\theta$\,$=$\,$180^{\degree}$. Furthermore, given that $\varepsilon_{r}$ is proportional to the magnon spin Nernst Hall conductivity $\alpha^{s}_{xy}$ in monolayer honeycomb antiferromagnets\,\cite{cheng2016spin, zyuzin2016magnon}, this observed dependence on the N\'{e}el vector may also manifest in magnon Hall transport phenomena. This intriguing coupling offers a unique approach for detecting N\'{e}el vector in 2D AFM insulators.   

\textit{Multilayer $\mbox{VP}X_{3}$.} Next, we examine the magnons of bilayer $\mbox{VP}X_{3}$, where the spin model can read as ${\cal \hat{H}}_{2}$\,$=$\,${\cal \hat{H}}_{1}$\,$+$\,$ {\cal \hat{H}}_{c}$, with ${\cal \hat{H}}_{c}$ representing AFM interlayer coupling term. Similar to the monolayer, the momentum space magnon Hamiltonian of bilayer ${\cal \hat{H}}_{2k}$ splits into two blocks. The first block, ${\cal \hat{H}}^{I}_{2k}$, is described by spinor basis $\psi^{\dagger}_{I,k} \equiv (\hat{a}_{1,k}^{\dagger},\hat{b}^{\dagger}_{2,k}, \hat{b}_{1,-k}, \hat{a}_{2,-k})$, where ${\hat{a}}^{\dagger}_{1(2),k}$ creates the magnon in first(second) layer at $k$-point. ${\cal \hat{H}}^{I}_{2k}$ read as  
\begin{equation}\label{H_2k}
    {\cal \hat{H}}^{I}_{2k} = \left(\lambda + g_{k}\right) I + 
    \begin{pmatrix} 
     f_{k}  & 0 & \gamma_{k} & {\cal J}_{c} \\
     0 & -f_{k} & {\cal J}_{c}& \gamma_{k}^{\dagger}\\
    \gamma_{k}^{\dagger} & {\cal J}_{c} & -f_{k}&  0 \\
    {\cal J}_{c} & \gamma_{k} & 0 & f_{k}\\
    \end{pmatrix}, 
\end{equation}
where ${\cal J}_{c}$\,$=$\,$0.6$\,$\mbox{meV}$ obtained by DFT calculations. The second block of Hamiltonian can be obtained by
$\gamma_{k}\rightarrow\gamma_{-k}$, $f_{k}\rightarrow f_{-k}$ in Eq.\,(\ref{H_2k}). By performing Bogoliubov transformation, the corresponding eigenvalues can be derived as
\begin{equation}\label{bafm_e}
\begin{split}
  \epsilon^{1}_{\pm} = \sqrt{\epsilon^{2}_{c} + \epsilon^{2}_{0}-2\sqrt{\left(\lambda + g_{k}\right)^{2}{f_{k}}^{2} - |\gamma_{k}|^{2} \epsilon^{2}_{c}}}, \\ 
  \epsilon^{2}_{\pm} = \sqrt{\epsilon^{2}_{c} + \epsilon^{2}_{0}+2\sqrt{\left(\lambda + g_{k}\right)^{2}{f_{k}}^{2} - |\gamma_{k}|^{2} \epsilon^{2}_{c}}} 
\end{split}
\end{equation}
with ${\epsilon^{2}_{c}}=f^{2}_{k}-{\cal J}^{2}_{c}$. 

\begin{figure}
\includegraphics[width=0.46\textwidth]{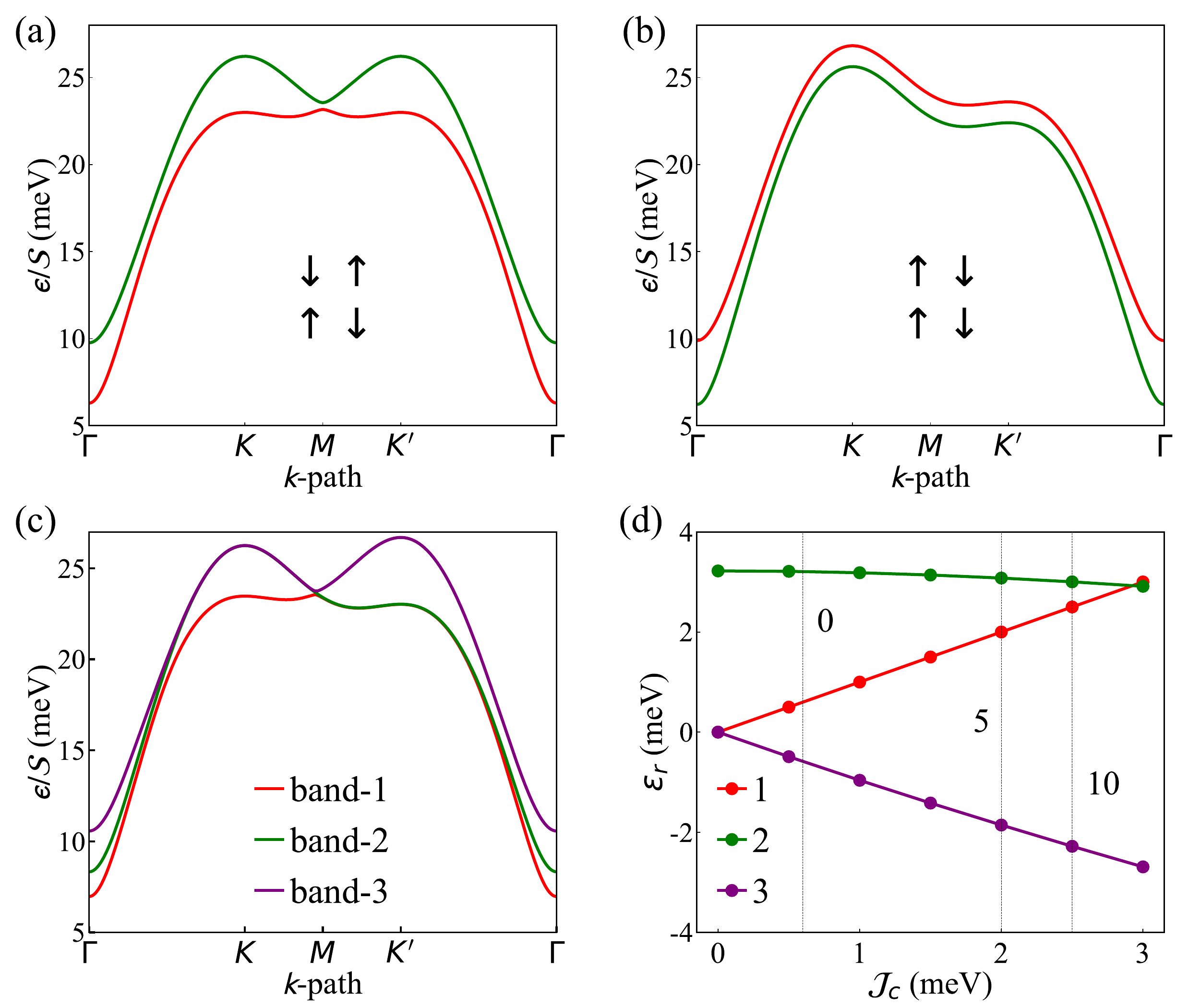}% Here is how to import EPS art
\caption{\label{fig4}(a) Magnon bands of bilayer $\mbox{VPTe}_{3}$ with ${\cal J}_{c}$\,$>$\,$0$. (b) ${\cal J}_{c}$\,$<$\,$0$. (c) The Magnon bands of trlayer $\mbox{VPTe}_{3}$. (d) The dependence of $\varepsilon_{r}$ for each band in trilayer $\mbox{VPTe}_{3}$ on interlayer coupling ${\cal J}_{c}$. The values of ${\cal J}_{c}$ under the applied the pressure 0\,Gpa, 5\,Gpa and 10\,Gpa are also labeled.}
\end{figure}

As shown in Fig.\,\ref{fig4}(a), the magnon bands of bilayer $\mbox{VP}{X}_{3}$ remains two-fold degenerate across the entire Brillouin zone, forming Dirac cones at $\mbox{K}$ points, which can be broken by the ${\cal D}_{z}$. Similar to the monolayer $\mbox{VP}{X}_{3}$, two-fold degenerate magnon modes carry opposite spin angular value $\langle{S^{z}}\rangle$ with chirality $\mp1$. However, in this bilayer, the corresponding magnon bands are symmetric, satisfying
$\epsilon(k)$\, $=$\,$\epsilon(-k)$, which indicates the absence of nonreciprocal magnons. This symmetric behavior is attributed to the
preserved ${\cal P}$ symmetry of magnetic order in bilayer $\mbox{VP}{X}_{3}$. Notably, this odd-even nature of layer dependence can be generalized to arbitrary even layers with AFM interlayer coupling. Therefore, beyond the 2NN DMI, the layer number\,($l_{n}$) plays a dominant role for nonreciprocal magnons in multi-layer honeycomb antiferromagnets.

\begin{figure}
\includegraphics[width=0.325\textwidth]{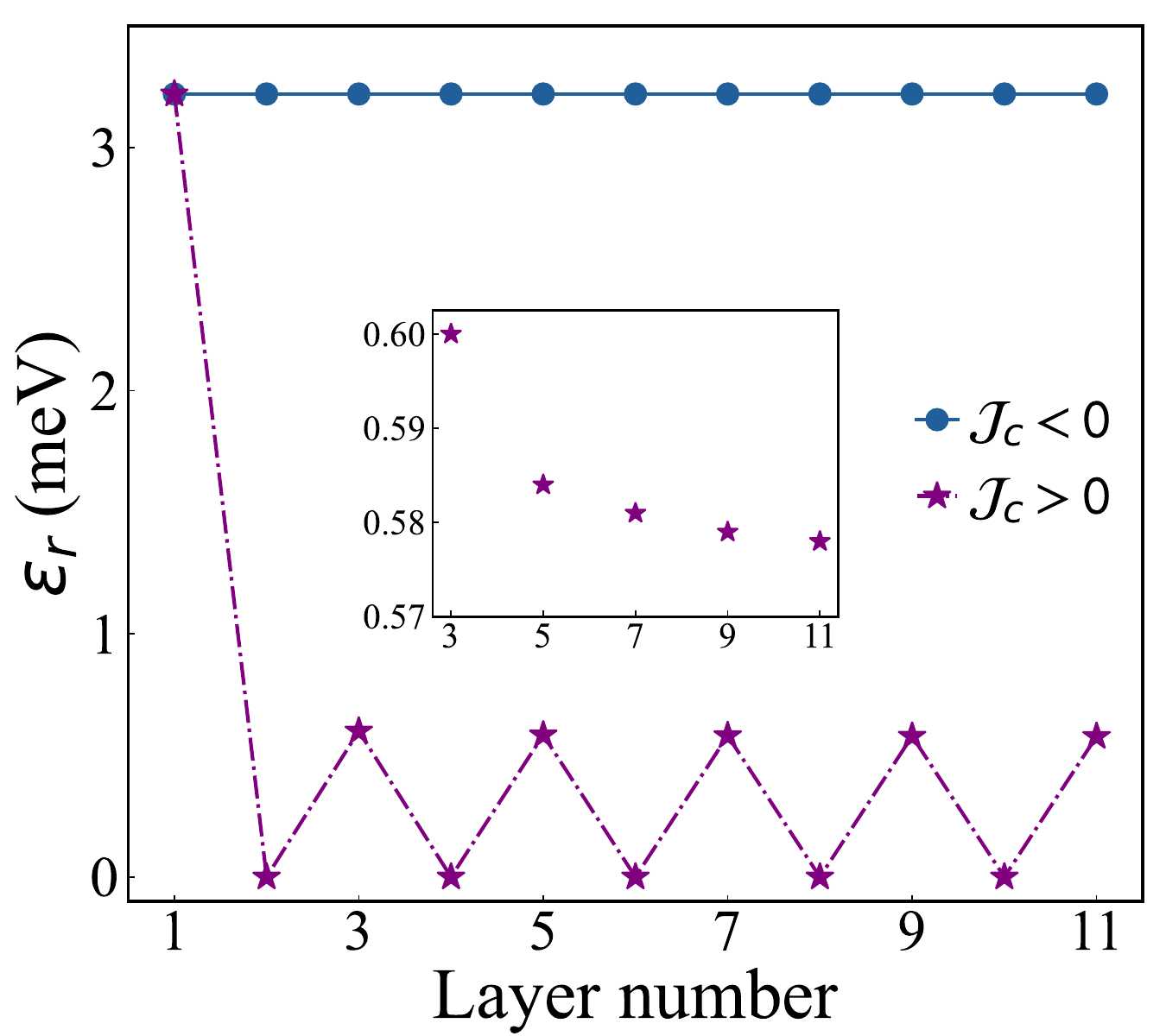}% Here is how to import EPS art
\caption{\label{fig5} The nonreciprocity of minor magnon band in layered $\mbox{VPTe}_{3}$ as a function of the layer number ($l_{n}$). Here, we compare the results for FM and AFM interlayer coupling.} 
\end{figure}

By contrast, in bilayers with ferromagnetic (FM) interlayer coupling (${\cal J}_{c}$\,$<$\,$0$), the ${\cal P}$ symmetry is broken. As shown in Fig.\,\ref{fig4}(b), both magnon bands exhibit nonreciprocal behavior. The minor band remains identical to that of the monolayer, whereas the major band is slightly shifted upward due to interlayer coupling. Fig.\,\ref{fig5} shows the dependence of magnon nonreciprocity in the minor band on the $l_{n}$ for multilayer honeycomb antiferromagnets with ferromagnetic ${\cal J}_{c}$. Notably, this behavior remains unchanged as $l_{n}$ increases.

Similarly, the magnon Hamiltonian of trilayer $\mbox{VP}X_{3}$ can be derived using the same approach. In trilayer,
the magnetic order breaks the ${\cal P}$ symmetry, which naturally leads to the presence of nonreciprocal magnons, irrespective of the type of interlayer coupling. As shown in Fig.\,\ref{fig4}(a), magnon band of trilayer $\mbox{VPTe}_{3}$ is two-fold degenerate and exhibits robust nonreciprocal magnons induced by ${\cal D}_{z}$. Interestingly, compared to the monolayer and bilayer, the trilayer exhibits a distinct difference in nonreciprocal magnons. Firstly, the direction of nonreciprocity differs among three magnon bands (each being two-fold degenerate), with
$\varepsilon_{r}$\,$>$\,$0$ for first band and $\varepsilon_{r}$\,$<$\,$0$ for second and third bands. Secondly, the magnitudes of ${\varepsilon}_{r}$ within each band are coupled to the interlayer coupling ${\cal J}_{c}$. More specifically, the $\varepsilon_{r}$ values of the first and third bands increase with ${\cal J}_{c}$ and an approximately linear relationship, as demonstrated by our numerical results shown in Fig.\,\ref{fig4}(d). In contrast, the $\varepsilon_{r}$ of the second band is suppressed by the ${\cal J}_{c}$. Given that the interlayer coupling can be readily tuned by the applied pressure or strain, it provides a new approach to manipulate the nonreciprocal magnons\,\cite{matsuoka2024mpx}. As shown in Fig.\,\ref{fig4}(d) and Fig.\,S4\,\cite{Supplemental_Materials}, our DFT calculations demonstrate that applying a pressure of only about $5\ \mbox{Gpa}$ is sufficient to increase ${\cal J}_{c}$ to $2\ \mbox{meV}$ in layered $\mbox{VPTe}_{3}$. Therefore, in addition to the DMI, interlayer coupling also plays a dominant role in nonreciprocal magnons in multilayer antiferromagnets and can be readily tuned by external fields.

Figure~\,\ref{fig5} illustrates the layer-dependent behavior of nonreciprocal magnons in layered $\mbox{VPTe}_{3}$. In addition to exhibiting an intriguing odd–even effect, the system also shows a clear trend of decreasing nonreciprocity with increasing layer number\,($l_{n}$). The value of $\varepsilon_{r}$ in the trilayer is only one-sixth of that in the monolayer, and follows $\varepsilon_{r}$\,$\propto$\,$1/\sqrt{l_{n}}$ when $l_{n}$ exceeds 3. This remarkable layer dependence reveals that nonreciprocal magnons in such vdW layered antiferromagnets exhibit fundamentally 2D behavior, providing a new insight for detecting their magnetic ground states. 

\textit{Summary and discussion.} 
According to the above symmetry analysis, breaking the inversion symmetry ${\cal P}$ of the magnetic structure is essential for realizing nonreciprocal magnons. This mechanism underpins the unique layer-dependent behavior observed in layered $\mbox{VP}X_{3}$. Indeed, beyond monolayer N\'{e}el type antiferromagnets, other magnetic configurations can also exhibit nonreciprocal magnons through stacking engineering. For example, in even-layer $\mbox{FeBr}_{3}$ or $\mbox{MnBi}_{2}\mbox{Te}_{4}$, while ferromagnetic order within each layer preserves ${\cal P}$ symmetry, the AFM AB-stacking antiferromagnetic arrangement breaks it\,\cite{lebegue2013two, cole2023extreme}. Given that the interlayer magnetic coupling in such vdW magnets can be tunable via external magnetic fields, magnon nonreciprocity provides a powerful probe for identifying their magnetic properties.

Recent studies have demonstrated that the bond-dependent spin exchanges can induce the nonreciprocal magnons\,\cite{gitgeatpong2017nonreciprocal, iguchi2015nonreciprocal}, for which $\langle{{\cal S}^{z}\rangle}$ no longer serves as a good quantum number. Our DFT calculations demonstrate that this interaction is weaker than 2NN ${\cal D}_{z}$ in monolayer $\mbox{VPX}_{3}$, with a magnitude of approximately $0.15\ \mbox{meV}$\,\cite{Supplemental_Materials}. This indicates that, in honeycomb antiferromagnets with N\'{e}el order, the dominant mechanism driving nonreciprocal magnons is 2NN DMI, and the resulting nonreciprocal magnons is characterized by the $\langle{{\cal S}^{z}\rangle}$. 

In summary, our work reports robust nonreciprocal magnons in layered antiferromagnets $\mbox{VP}{X}_{3}$. The revealed N\'{e}el order dependence offers a unique avenue for detecting the AFM order parameters in the 2D limit. Moreover, in multilayers, nonreciprocal magnons exhibit intriguing layer-dependent behavior combined with high tunability, underscoring their promising potential for future wave-based, ultra-compact spintronic applications.

This work is supported by the National Natural Science Foundation of China (Grant No. 12374092), Natural Science Basic Research Program of Shaanxi (Program No. 2023-JC-YB-017), Shaanxi Fundamental Science Research Project for Mathematics and Physics (Grant No. 22JSQ013), “Young Talent Support Plan” of Xi'an Jiaotong University, the Open Project of State Key Laboratory of Surface Physics (Grant No.\ KF2023\_06), and the Xiaomi Young Talents Program. L. B. thanks the Vannevar Bush Faculty Fellowship Grant No. N00014-20-1C2834 from the Department of Defense and Grant No. DMR-1906383 from the National Science Foundation Q-AMASE-i Program (MonArk NSF Quantum Foundry) 

\textit{Data availability}. All data are available from the authors upon reasonable request.

\appendix
%\begin{verbatim}
%\end{verbatim}

\nocite{*}
%\nobalance
%\bibliographystyle{plain}
%\bibliographystyle{abbrv}
%\bibliographystyle{elsarticle-num-names}
%\bibliographystyle{apsrev4-2}
\bibliography{Main}% Produces the bibliography via BibTeX.

\end{document}